\documentclass[conference]{IEEEtran}

\usepackage{graphicx} 
\usepackage{amsmath} 
\usepackage{color,soul} 
\usepackage{xcolor}
\usepackage{amssymb} 
\usepackage{textcomp} 
\DeclareUnicodeCharacter{2212}{\textminus}
\usepackage{algorithm} 
\usepackage{cite} 
\usepackage{booktabs} 
\usepackage{subcaption} 
\usepackage{url}
\usepackage{enumitem}

\usepackage{mathtools}

\usepackage{algpseudocode} 

\usepackage{caption}
\DeclareMathOperator{\cvar}{CVaR}
\DeclareMathOperator{\var}{VaR}
\DeclareMathOperator{\poe}{PoE}
\DeclareMathOperator{\bpoe}{BPoE}

\title{Managing Risk using Rolling Forecasts in Energy-Limited and Stochastic Energy Systems}
\author{Thomas Mortimer, Robert Mieth}
\date{March 2024}

\usepackage{placeins}
\newcommand{\subparagraph}{}
\usepackage{titlesec}
\titlespacing*{\subsection}{0pt}{4pt}{0pt}
\begin{document}

\bstctlcite{IEEE:BSTcontrol} 

\maketitle

\label{sec:abstract}

\textbf{\textit{Abstract}---
We study risk-aware linear policy approximations for the optimal operation of an energy system with stochastic wind power, storage, and limited fuel.
The resulting problem is a sequential decision-making problem with rolling forecasts. 
In addition to a risk-neutral objective, this paper formulates two risk-aware objectives that control the conditional value-at-risk of system cost and the buffered probability of exceeding a predefined threshold of unserved load.
The resulting policy uses a parameter-modified cost function approximation that reduces the computational load compared to the direct inclusion of those risk measures in the problem objective.
We demonstrate our method on a numerical case study.
}
\section{Introduction} \label{sec:introduction}
   
In energy systems that largely rely on electric power generation from wind and solar, security of supply cannot be ensured only by matching installed generation capacity with peak demand plus some reserve \cite{Stenclik2021}.
Instead, system operators require new sets of tools alongside synergistic energy resources (e.g., battery or power-to-X storage technology) that can translate variable and uncertain renewable power production into reliable and continuous supply of demand \cite{Yankson2023}.
This challenge is amplified in energy-limited systems where falling back to a virtually infinite source of fuel-based generation \cite{taylor2016power} or supply from a central grid is not available \cite{hans2019risk}. 
As a result, system operators require risk-aware decision-making tools that guarantee pre-defined reliability targets (e.g., Loss-of-Load Probability--LOLP) while remaining cost-efficient.
Resulting operational decisions on controllable resources, e.g., charging or discharging storage, using available fuel, or calling upon flexible demand resources, depend on the real-time production from wind and solar, forecasts of their production, and the statistical properties of these forecasts. 
The energy- and use-limited nature of these controllable resources in combination with wind and solar forecast uncertainty creates a complex sequential decision-making problem that quickly becomes intractable in practical settings \cite{Powell2011}. 
Moreover, it is not straightforward to translate the day-to-day usage of the portfolio of available controllable resources into effective capacity values that can be used for system planning purposes \cite{Burdick2022}.

This paper takes a step towards addressing these challenges building on results by Ghadimi and Powell \cite{Ghadimi2022} who present a parameter-modified cost function approximation for the sequential decision-making problem of an energy storage model with rolling forecasts.
In particular, we modify the model in \cite{Ghadimi2022} such that we can directly manage system risk by enforcing Conditional Value-at-Risk ($\cvar$) or Buffered Probability-of-Exceedance ($\bpoe$) constraints. This allows the system operator to directly enforce pre-defined reliability targets. 
The resulting decision policy takes the form of a deterministic look-ahead model with an offline-tuned discount parameter that offers insights on how wind and solar production time series should be adjusted in planning simulations to reflect the reality of system operations.

\subsection{Related Literature}

Besides \cite{Ghadimi2022}, parameter-modified cost function approximation for an energy storage model has been investigated in \cite{Shuai2019,Wang2023}, which propose alternative tuning strategies and do not include risk-awareness.
In \cite{Shuai2019} an online learning strategy is used instead of training the parameter in a simulator and in \cite{Wang2023} a transfer function is used to obtain the parameter values.
Managing risk in energy-limited and stochastic energy systems has been studied extensively, e.g., in \cite{Nasr2019risk-averse_microgrid,Yankson2023,Bakhtiari2022}. This paper differs from the other studies by including a rolling forecast and formulating the system as a sequential decision-making problem, enabling the possibility of investigating the evolution of the forecast over time. 
Other studies on risk-averse sequential decision-making in energy systems have been introduced in the form of risk-averse model predictive control (MPC) methods, e.g., in \cite{hans2019risk,Ning2021,Mark2022,Khodabakhsh2016}.
Relative to these works that manage risk using $\cvar$, this paper uses the cost function approximation approach from \cite{Ghadimi2022} and enforces BPoE as an additional means of managing risk on technical constraints.
Managing risk using $\bpoe$ has been highlighted in other engineering and financial domains \cite{Rockafellar2010,Mafusalov2018}.

\section{Energy System Model} \label{sec:energy_storage_model}

\begin{figure}
    \centering
    \includegraphics[width=0.95\linewidth, keepaspectratio]{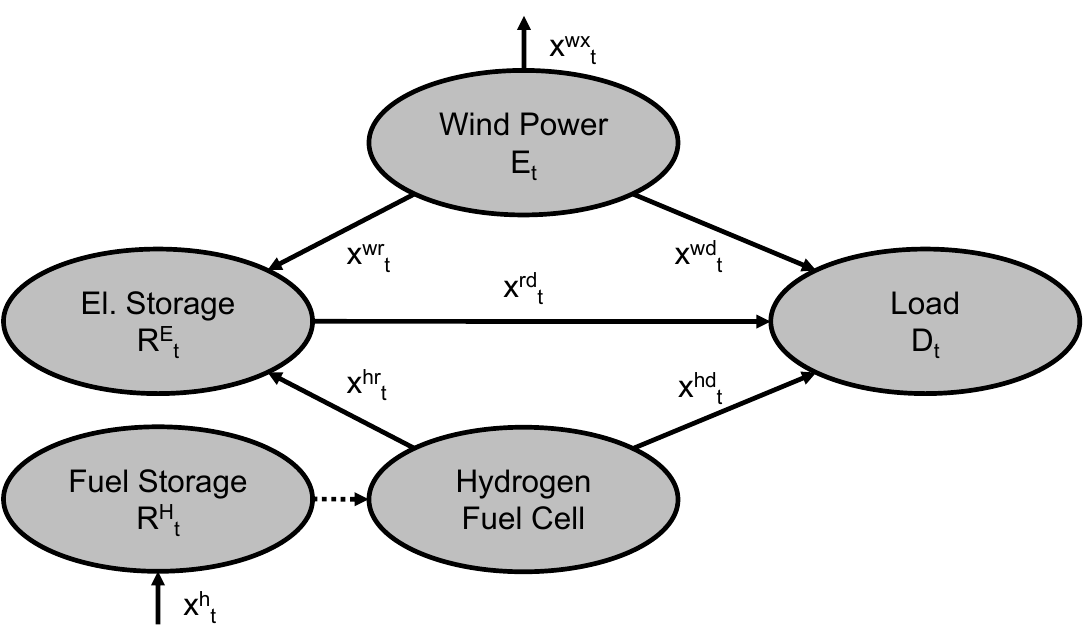}
    \caption{Energy system model with power exchange between system components as decision variables. Arrows indicate variable signs.}
    \label{fig:model_overview}
\end{figure}

We study a risk-averse decision-maker that continuously operates an energy system with uncertain power injections from a wind generator and fluctuating energy and fuel prices.
Figure~\ref{fig:model_overview} shows a schematic of the energy system model.

The system operator manages stochastic and energy-limited resources with the objective of meeting electricity demand in a cost-effective and reliable manner.
We model the decision-making process through discrete time steps indexed by $t$ over an operational time horizon of $T$ steps.
The main energy source is wind power.
In addition, the operator has access to an electric battery system, fuel-based generation, and fuel storage. 
We assume that fuel can only be acquired at certain time steps forcing the operator to plan and purchase the required amount for current and future time periods. 
This can for example represent an islanded microgrid with a gas or hydrogen storage that receives fuel shipments at a set schedule \cite{Zia2018}. 
In this paper we assume hydrogen powering a fuel cell but note that this is not required for the model.

For each $t$ we define the vector of decision variables as
\mbox{$x_t = (x_t^{wd}, x_t^{rd}, x_t^{hd}, x_t^{wr}, x_t^{hr}, x_t^{h}, x_t^{wx})^{\top}$}, where:
\begin{itemize}
    \item $x_t^{wd}$: Energy from wind satisfying load;
    \item $x_t^{wr}$: Energy from wind saved in the battery;
    \item $x_t^{wx}$: Curtailed wind energy. 
    \item $x_t^{rd}$: Energy from battery satisfying load;
    \item $x_t^{hd}$: Energy from the fuel cell satisfying the load;
    \item $x_t^{hr}$: Energy from the fuel cell stored in the battery;
    \item $x_t^{h}$: Amount of hydrogen purchased. 
\end{itemize}
Further, in each time step $t$ decision $x_t$ is constrained by:
\allowdisplaybreaks
\begin{subequations}
\begin{align}
    && x_t^{wr} + x_t^{wd} + x_t^{wx} &\leq E_t  \label{eq:Wind_Power_Limit_constraint}  \\
    && x_{t}^{wd} + \beta^d x_t^{rd} + \beta^h x_t^{hd} &\leq D_t  &\label{eq:power_balance} \\
    && x_t^{h} &\leq T^H_t R^{\bar{H}}   \label{eq:Fuel_Day_acquisition_constraint}  \\
    && x_t^{rd} &\leq R^E_t  \label{eq:Battery_Storage_SoC_constraint}  \\
    && x_t^{hr} + x_t^{hd} &\leq R^H_t  \label{eq:Fuel_Storage_SoC_constraint}  \\
    && x_t^{h}  &\leq R^{\bar{H}} - R^H_t \label{eq:Fuel_Storage_Charge_limit_by_SoC_constraint}  \\    
    && \beta^c (x_t^{wr} + \beta^h x_t^{hr}) - x_t^{rd}   &\leq R^{\bar{E}}  - R^E_t \label{eq:Battery_Storage_Charge_limit_by_SoC_constraint}  \\
    && x_t^{wr} + \beta^h x_t^{hr} &\leq \gamma^c  \label{eq:Battery_Storage_Charge_limit_by_Charging_capacity_constraint} \\
    && x_t^{rd} &\leq \gamma^d  \label{eq:Battery_Storage_Discharge_limit_by_Discharging_capacity_constraint}  \\
    && \beta^h( x_t^{hr} + x_t^{hd}) &\leq \gamma^h \label{eq:Fuel_Storage_Discharge_limit_by_Discharging_capacity_constraint} \\
    && R^H_t - x_t^{hd} - x_t^{hr} + x_t^{h}  &=  R^H_{t+1} \label{eq:Fuel_transition_function}  \\
    && R^E_t - x_t^{rd} + \beta^c (x_t^{wr} + x_t^{hr}) &= R^E_{t+1} \label{eq:Battery_storage_transition_function} \\
    && x_t &\geq 0. \label{eq:non-negativity} 
\end{align}
\label{eq:energy_storage_constraints}%
\end{subequations}%
\allowdisplaybreaks[0]%
Constraint~\eqref{eq:power_balance} relates demand $D_t$ to production from wind, the battery, and the fuel cell. 
Constraints~\eqref{eq:Battery_Storage_SoC_constraint} and \eqref{eq:Battery_Storage_Discharge_limit_by_Discharging_capacity_constraint} limit power production from the battery by the current state of charge $R_t^E$ and the battery power rating $\gamma^d$, respectively.
Constraints~\eqref{eq:Fuel_Storage_SoC_constraint} and \eqref{eq:Fuel_Storage_Discharge_limit_by_Discharging_capacity_constraint} limit the total power production of the fuel cell to the available energy stored as hydrogen $R_t^H$ and its power rating $\gamma^h$, respectively subject to the fuel cell efficiency. 
Constraint~\eqref{eq:Wind_Power_Limit_constraint} limits the total power production from wind to the currently available wind power $E_t$.
Hydrogen acquisition is limited by the remaining available storage capacity in Constraint~\eqref{eq:Fuel_Storage_Charge_limit_by_SoC_constraint}. 
In Constraint~\eqref{eq:Fuel_Day_acquisition_constraint} hydrogen acquisition is limited to a subset of timesteps defined by $T^H\in\{0,1\}^T$, i.e., $T^H_t = 1$ if hydrogen can be acquired in time step $t$ and $0$ otherwise.
Battery charging and discharging is subject to the battery efficiency $\beta^c$ and limited by remaining available storage capacity in Constraint~\eqref{eq:Battery_Storage_Charge_limit_by_SoC_constraint} and the battery power rating in Constraint~\eqref{eq:Battery_Storage_Charge_limit_by_Charging_capacity_constraint}.
Charging of the battery storage is also limited by the charging capacity in Constraint~\eqref{eq:Battery_Storage_Discharge_limit_by_Discharging_capacity_constraint}. 
Constraint~\eqref{eq:non-negativity} enforces non-negativity for all decision variables.
Constraints~\eqref{eq:Battery_storage_transition_function} and \eqref{eq:Fuel_transition_function} are the time-coupled storage constraints.

We define $S_t$ as the state variable containing all the information needed to solve a problem objective with respect to constraints~\eqref{eq:energy_storage_constraints}.
The initial state at $t=0$ is predefined as $S_0$. 
At each $t$, state $S_t$ is defined by:
\begin{itemize}
    \item $D_t$: Electricity demand at time $t$.
    \item $E_{t}$: Wind power at time $t$
    \item $P^H_t$: Cost of hydrogen at time $t$.
    \item $R^E_t$: The level of energy in the battery satisfying $R^E_t \in [0, R^{\bar{E}}]$, where $R^{\bar{E}} > 0$ represents the capacity.
    \item $R^H_t$: Level of energy in the hydrogen storage satisfying $R^H_t \in [0, R^{\bar{H}}]$, where \mbox{$R^{\bar{H}} > 0 $} represents the capacity. 
\end{itemize}
Additional parameters are:
\begin{itemize}
    \item $\beta^c$: Battery charging efficiency.
    \item $\beta^d$: Battery discharging efficiency.
    \item $\beta^h$: Fuel cell generation efficiency.
    \item $\gamma^c$: Battery charging limit. 
    \item $\gamma^d$: Battery discharging limit.
    \item $\gamma^h$: Fuel cell generation limit.
    \item $C^P$: Penalty cost for unsatisfied load.
    \item $C^W$: Penalty cost for curtailed wind.
\end{itemize}
Available wind power production is uncertain for all timesteps $\{t+1, T\}$, demand and hydrogen are assumed to be known for ease of exposition in this paper. 
Power demand and hydrogen prices are assumed to be known over the time horizon. 

The system operator minimizes its cost function:
\begin{equation}
    C_t(\!S_t,x_t)\!\! =\! C^P\! L_t(S_t,x_t)\!+\!C^W x_t^{wx}\!\!+ P_t^h x_{t}^{h},\label{eq:cost_function}
\end{equation}
where mismatch between demand and supply is given by
\begin{equation}
        L_t(S_t,x_t) = D_t - (x^{wd}_t + \beta^d x^{rd}_t + \beta^h x^{hd}_t). \label{Loss_function}
\end{equation}
Each unit of demand mismatch $L_t(S_t,x_t)$ incurs cost $C^P$. (Note the non-negativity condition \eqref{eq:power_balance}.)
Cost $C^{W}$ penalizes wind curtailment and the final term in \eqref{eq:cost_function} captures the cost of purchasing fuel $x_{t}^{h}$ at price~$P_t^h$.

\section{Risk-aware decision-making} \label{sec:risk-aware_decisions}
For any future $t'>t$, cost function \eqref{eq:cost_function} depends on uncertain system states $S_t(\omega)$ where $\omega\in\Omega$ is a random variable. To account for this uncertainty, decision-makers instead minimize expected cost $\mathbb{E}\big[\sum_{t'=t}^T(C_{t'}(S_{t'}(\omega), x_{t'}))\big],\ \forall t'$, e.g., as in \cite{Ghadimi2022}. 
This risk-neutral approach, however, is often not suitable for practical applications where decision-makers are risk-averse. This section briefly discusses options to handle risk.

\subsection{Risk Measure: Conditional Value-at-Risk} 
\label{subsec:cvar}
Conditional Value-at-Risk ($\cvar$) is a popular risk metric in energy modelling due to its convex properties \cite{Stenclik2024}.
$\cvar$ is defined via the Value-at-Risk (VaR) $q_{\alpha}(X)$, which, for a given $\alpha \in [0,1]$, returns the minimum value $z$ that a random variable $X$ will not exceed with probability $\alpha$:
\begin{align}
    \var_{\alpha} (X):= q_{\alpha}(X) = \min \Big\{z \mid \mathbb{P}(X \leq d) \geq \alpha \Big\}. \label{eq:VaR}
\end{align}
VaR has useful applications for chance-constrained programs under some conditions, e.g., \cite{bienstock2014chance,lubin2015robust,mieth2019distribution}, but typically results in non-convex formulations.

$\cvar$ ($\bar{q}_{\alpha}$), on the other hand, captures the expected value of $X$ \textit{under the condition} that $X$ exceeds $q_{\alpha}$ \cite{Mafusalov2018}:
\begin{align}
  \cvar_{\alpha}(X) := \bar{q}_{\alpha}(X) = \mathbb{E} \Big[ X \mid X > q_{\alpha}(X) \Big]. \label{function:CVAR}
\end{align} 
For a linear cost function and assuming discrete outcomes (scenarios) $X_{\omega}$, $\cvar$ $\bar{q}_{\alpha}(X)$ can be minimized in a tractable linear program \cite{Mafusalov2018}:
\begin{subequations}
\begin{align}
    &\min_{x,z \in \mathbb{R},y_{\omega} \geq 0}
    && z + \frac{1}{1-\alpha } \sum_{\omega = 1}^{\Omega} \frac{1}{|\Omega|} y_{\omega}  \label{cvar_obj}  \\ 
    & \text{s.t.} && C(X_{\omega},x) - z - y_{\omega} \leq 0 & \forall \omega \in \Omega \label{cvar_const_1}.
\end{align}   
\label{eq:cvar_opt}
\end{subequations}

\begin{figure}
    \centering
    \includegraphics[width=0.4\textwidth, keepaspectratio]{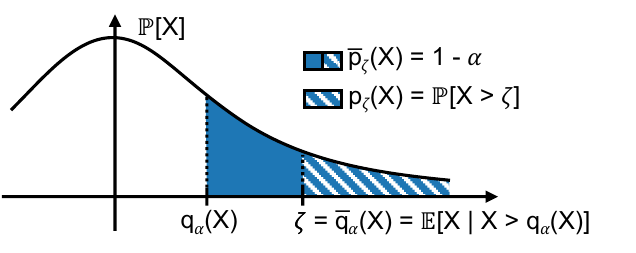}
    \caption{Probability density function of a continuous random variable $X$. For a threshold $\zeta \in \mathbb{R}$, $p_x(X)$ equals $\mathbb{P}(X > z)$, which is the cumulative density (hashed area). For the same threshold $\zeta$, $\bar{p}_{\zeta}(X)$ is the cumulative density (solid+hashed area). The expectation of the worst-case $1 - \alpha = \bar{p}_{\zeta}(X)$ outcomes equals $\zeta = \bar{q}_{\alpha}(X)$.
    }
    \label{fig:CVAR_and_BPOE}
\end{figure}
Fig.~\ref{fig:CVAR_and_BPOE} illustrates the relation between $q_{\alpha}$ and $\bar{q}_\alpha$.
$\cvar$ allows to minimize the expected cost of the $100\alpha$-percent worst cases. In many engineering applications, however, risk-targets are not defined in terms of severity (probability times outcome), but whether a security threshold is maintained with high probability. 
We discuss this in the following section.

\subsection{Risk Measure: Buffered Probability-of-Exceedance} \label{subsec:bpoe}

Analogous to the relationship between VaR and $\cvar$, Buffered Probability-of-Exceedance ($\bpoe$) is defined via the Probability-of-Exceedance (PoE, or reliability), which
quantifies the probability mass of a random variable $X$ exceeding a given threshold $\zeta \in \mathbb{R}$ (see also Fig.~\ref{fig:CVAR_and_BPOE}):
\begin{align}
    \poe_{\zeta} := p_{\zeta}(X) = \mathbb{P}\Big[X > \zeta \Big].
\end{align}

PoE suffers the same mathematical shortcomings as VaR.
As the counterpart to $\cvar$ , $\bpoe$ overcomes these shortcomings and 
computes the risk-level $\alpha$ at which the $\cvar$ is equal to the predefined threshold $\zeta$ \cite{Mafusalov2018}:
\begin{align}
    \bar{p}_{\zeta}(X) = \underset{\gamma \geq 0}{\min} \quad \mathbb{E}\Big[\gamma ( X - \zeta) + 1 \Big]^{+},\label{eq:bpoe} 
\end{align} 
where $\gamma$ is an auxiliary variable and $[\cdot]^+ = \max\{\cdot, 0\}$.
Similarly to $\cvar$, minimizing $\bpoe$ over a set of scenarios $X_{\omega},\ \omega\in\Omega$ can be written as the linear program \cite{Mafusalov2018}:
\begin{subequations}
    \begin{align}
    &\underset{x, \gamma \geq 0 , \eta_{\omega} \geq 0}{\min} \ && 
    \sum_{\omega = 1}^{\Omega} \frac{1}{\Omega} \eta_{\omega} \label{bpoe_obj}  \\ 
    & \text{s.t.} && \gamma C(X_{\omega},x)\! - \!\gamma \zeta \!+\! 1\! -\! \eta_{\omega} \!\leq\! 0 & \forall \omega \in \!\Omega.  \label{bpoe_const}
    \end{align}%
    \label{bpoe_reformulation}%
\end{subequations}%
Term $\gamma C(X_{\omega},x)$ in \eqref{bpoe_const} allows the convex reformulation \cite{Mafusalov2018}:
\begin{equation}
    (C\gamma)(x) =  \begin{cases}
      \gamma C(x/\gamma) & \gamma > 0 \\
      0 &  \gamma = 0, x = 0 \\
      + \infty & \gamma = 0, x \neq 0.
    \end{cases}
\end{equation}

In contrast to $\cvar$, which has been connected to energy-based reliability metrics such as expected energy-not-served (EENS), e.g., in \cite{mieth2022risk}, the properties of $\bpoe$ allow to incorporate predefined frequency-based reliability metrics, e.g., the ``1 day in 10 years'' rule \cite{nerc2022Reexamining} or loss-of-load probability.
Hence, for a given $\zeta$
$\bpoe$ returns the probability that $C$ does not exceed $\zeta$ for a given decision $x$, thus \textit{certifying reliability.} 

\section{Risk-aware policies for energy system model} \label{sec:policy}

\begin{figure}
    \centering
    \includegraphics[width=0.4\textwidth, keepaspectratio]{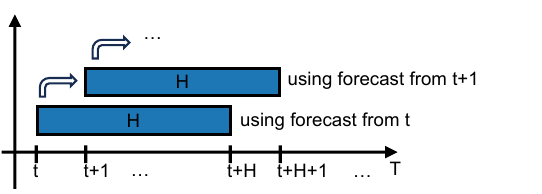}
    \caption{Rolling horizon illustration}
    \label{fig:Rolling_Horizon}
\end{figure}

We now seek a risk-aware decision policy to solve the energy model from Section~\ref{sec:energy_storage_model} as a sequential decision-making problem with a rolling wind power forecast. 
We define a policy as a function $X(S_t)$ that returns a decision $x_t$ given the current state $S_t$. 
Considering a rolling wind power forecast allows us to model a realistic decision-making process in which the operator only needs to commit to here-and-now decisions and can adjust look-ahead decisions with access to more accurate forecasts in the next time step.
We also model a limited operating horizon $H\le T$ as it can be ineffective to make decisions for time periods too far into the future.
Fig.~\ref{fig:Rolling_Horizon} illustrates the rolling forecast and horizon $H$.

\subsection{Stochastic risk-aware policies} \label{subsec:policy_design_stoch_risk_opt}

As a baseline, we consider a scenario-based risk-neutral look-ahead strategy (S-LA). 
The S-LA policy computes decisions that minimize cost at the current time step $t$ and the expected cost for future time steps $t<t'\le t+H$ over a set of scenarios $\omega \in \Omega$.
We denote all scenario-dependent variables with scenario index $\omega$.
The resulting S-LA policy is:
\allowdisplaybreaks
\begin{subequations}
\begin{align}
    &X_t^{\rm S-LA}(S_t): \notag \\
    \min_{x_t,x_{\omega t}}&  C_t(S_t,x_t)  + \frac{1}{|\Omega|}  \sum^{\Omega}_{\omega = 1} \sum^{\min(t+H,T)}_{t' = t+1} C_{t'}(S_{t'}(\omega),x_{t'})  \\
    \text{s.t.} & \quad  t: \eqref{eq:energy_storage_constraints} \notag \\
    &\forall t':\eqref{eq:Battery_Storage_SoC_constraint}-\eqref{eq:non-negativity}, \notag  \\
    &x_{\omega t'}^{wd} + \beta^d x_{t'}^{rd} + x_{t'}^{hd}   \leq D_{t'} \quad \forall \omega \in \Omega \label{eq:power_balance_LA}  \\  
    &x_{t'}^{wr} + x_{\omega t'}^{wd} + x_{\omega t'}^{wx}  \leq f^E_{\omega t'}  \quad \forall \omega \in \Omega \label{eq:Wind_Power_Limit_constraint_LA} \\
    &x_{t'}^{h}  \leq H^H_t R^{\bar{H}},   \label{eq:Fuel_Day_acquisition_constraint_LA} 
\end{align}%
\label{eq:Stochastic_Look_Ahead_model}%
\end{subequations}%
\allowdisplaybreaks[0]%
where $f^E_{\omega t'}$ is the available wind power at time $t'$ in scenario $\omega$ (e.g., obtained from a probabilistic forecast). 
Constraints \eqref{eq:power_balance_LA}, \eqref{eq:Wind_Power_Limit_constraint_LA} are the scenario-dependent counter-parts of \eqref{eq:power_balance}, \eqref{eq:Wind_Power_Limit_constraint}.
Constraint \eqref{eq:Fuel_Day_acquisition_constraint_LA} alters \eqref{eq:Fuel_Day_acquisition_constraint} such that only the time steps that allow hydrogen purchase within the horizon are considered.

Next, we modify the risk-neutral S-LA strategy to become risk-aware. 
The resulting risk-averse stochastic look-ahead (S-CVaR) approach computes decisions that minimize the expected cost in the $100\alpha$-percent worst case scenarios using CVaR (see Section~\ref{sec:risk-aware_decisions}):
\allowdisplaybreaks
\begin{subequations}
\begin{align}
    X_t^{\rm S-CVaR}(S_t): \notag & \\
    \min_{x_t,x_{\omega t},y_{\omega},z}& C_t(S_t,x_t) +  z + \frac{1}{1 - \alpha} \sum^{\Omega}_{\omega = 1} \frac{1}{|\Omega|}y_{\omega} \label{eq:Cvar_reformulation_objective}
\end{align} 
\vspace{-0.5cm}
\begin{align}
    \text{s.t.} \quad t:&  \eqref{eq:energy_storage_constraints} \notag\\
    \forall t':&\eqref{eq:Battery_Storage_SoC_constraint}-\eqref{eq:non-negativity},\eqref{eq:power_balance_LA}-\eqref{eq:Fuel_Day_acquisition_constraint_LA} \notag \\
    \sum^{\min(t+H,T)}_{t' = t+1}& C_{t'}(S_{t'}(\omega),x_{t'}) - z - y_{\omega} \leq 0 & \forall \omega \in \Omega \label{eq:Cvar_reformulation_constraint} \\
    & y_{\omega}  \geq 0 & \forall \omega \in \Omega, \label{eq:non-negativity_CVAR} 
\end{align}%
\label{eq:CVAR_Stochastic_Look_Ahead_model}%
\end{subequations}%
\allowdisplaybreaks[0]%
where we use auxiliary variables $z$, $y_{\omega}$ for the CVaR representation from \eqref{eq:cvar_opt}. Note that $\alpha$ is a parameter in S-CVaR that must be pre-defined by the system operator.

Instead of defining $\alpha$ beforehand, an operator might be more interested in ensuring that a pre-defined load-not-served threshold is maintained with maximum probability. 
We define this threshold as $\gamma$.
In this approach we minimize expected cost as in S-LA and additionally minimize $\bpoe$ of $\sum_{t`=t+1}^{\min(t+H,T)}L_{t`}$ with respect to $\gamma$. 
We denote this approach S-BPoE given by:
\allowdisplaybreaks
\begin{subequations}
\begin{align}
    X_t^{\rm S-BPoE}(S_t):& \notag \\
    \min_{x_t,x_{\omega t},\eta_{\omega},\zeta}& C_t(S_t,x_t) + \frac{1}{|\Omega|}  \sum^{\Omega}_{\omega = 1} \sum^{\min(t+H,T)}_{t' = t+1} C_{t'}(S_{t'}(\omega),x_{t'}) \notag \\ 
    &+M \frac{1}{|\Omega|}\sum^{\Omega}_{\omega = 1} \eta_{\omega}
    \label{eq:BPoE_reformulation_objective}
\end{align}
\allowdisplaybreaks
\begin{align}
    \text{s.t.} &\quad t: \eqref{eq:energy_storage_constraints} \notag\\
    &\forall t':\eqref{eq:Battery_Storage_SoC_constraint}-\eqref{eq:non-negativity},\eqref{eq:power_balance_LA}-\eqref{eq:Fuel_Day_acquisition_constraint_LA},& \notag \\
    &\gamma (\!\sum^{\min(t+H,T)}_{t'\!= t+1}\!L_{t'}(S_{t'}(\omega),x_{t'})) - \gamma \zeta + 1  \leq  0 & \forall \omega \in \Omega \label{eq:BPoE_reformulation_constraint} \\
    &\eta_{\omega}   \geq 0 & \forall \omega \in \Omega \label{eq:non-negativity_BPOE} 
\end{align}%
\label{eq:BPoE_Stochastic_Look_Ahead_model}%
\end{subequations}%
\allowdisplaybreaks[0]%
We use a large scalar $M$ to ensure that meeting the reliability target is prioritized over cost minimization.

For large-scale problems and/or large numbers of scenarios, the stochastic approaches S-LA, S-CVaR, and S-BPoE can become computationally intractable. 
The following section introduces a parameter-modified cost function approximation that can still capture risk aversion but
shifts computational load from decision-making to tuning the parameter. 

\subsection{Parameter-modified cost function approximation policies} \label{subsec:policy_design}

The goal of this subsection is to find a \textit{single} wind power scenario equivalent that performs well compared to the stochastic approaches from Section~\ref{subsec:policy_design_stoch_risk_opt} above.
We write this deterministic look-ahead (D-LA) model as:
\allowdisplaybreaks
\begin{subequations}
\begin{align}
    X_t^{\rm D-LA}(S_t):& \notag \\
    \min_{x_t} \quad & C_t(S_t,x_t)  +  \sum^{\min(t+H,T)}_{t' = t+1} C_{t'}(S_{t'}(\omega),x_{t'})  \\
    \text{s.t.}&  \quad t: \eqref{eq:energy_storage_constraints} \notag \\ 
    &\forall t': \eqref{eq:power_balance}-\eqref{eq:non-negativity}, \notag \\ 
    &x_{t'}^{wr} + x_{t'}^{wd} + x_{t'}^{wx}   \leq b_{t'}(f^E_{t'}, \theta) \label{eq:Parameter_Modification_DLA} 
\end{align}
\label{eq:DLA_model}
\end{subequations}
\allowdisplaybreaks[0]
Model D-LA includes an additional parameter $\theta$ that modifies the wind power scenario $f_{t'}^E$ to the upper bound $b_{t'}(f_{t'}^E, \theta)$ in \eqref{eq:Parameter_Modification_DLA}.
Modification $b_{t'}(f_{t'}^E, \theta)$ can be chosen as:

\noindent \textbf{Constant}: $b_{t'}(f^E_{t'}, \theta) = \theta f^E_{t'} $ All future forecast values are equally discounted.
    
\noindent \textbf{Look-up table}: $b_{t'}(f^E_{t'}, \theta) =  \theta_{t'-t}  f^E_{t'}$ with a different $\theta_{\tau}$ for each look-ahead period $\tau = 0,1,2,...$ with $\tau = t'-t$. 
    All future wind forecast values are discounted with an individual parameter depending on the look-ahead distance.

Parameter $\theta$ needs to be tuned offline as we describe below. Choosing a single ``Constant'' parameter will reduce tuning effort and further simplify the model. The ``Look-up table'' approach, on the other hand, increases modelling fidelity but also computational efforts in both tuning and solving the model.
We also refer to \cite{Ghadimi2022} for more discussion.

\subsection{Parameter tuning}

Our goal is to tune $\theta$ such that the resulting deterministic policy in D-LA achieves the goals of the risk-averse stochastic programs discussed in Section~\ref{subsec:policy_design_stoch_risk_opt}.
We formalize the resulting parameter tuning problems corresponding to S-LA, S-CVaR, and S-BPoE, respectively, as:

\noindent\textbf{Expected cost}: 
    Parameter $\theta$ is tuned such that \eqref{eq:DLA_model} minimizes expected cost.
    \allowdisplaybreaks
    \begin{align}
        & \underset{\theta}{\min} \Bigg\{ F^{EC}(\theta):= \mathbb{E}_{\omega} \Big[ F(\theta,\omega) \Big] \notag \\
        & = \mathbb{E} \Big[ \sum_{t=0}^{T} C_t(S_t(\omega),X^{\rm D-LA}_t(S_t(\omega) \big| \theta)) \Big| S_0 \Big] \Bigg\}. \label{eq:EX_tuning_problem}  
    \end{align}
    
\noindent \textbf{Risk-aware cost}: 
    Parameter $\theta$ is tuned such that \eqref{eq:DLA_model} minimize the 100$\alpha$-percent worst-case cost outcomes:
    \allowdisplaybreaks
    \begin{align}
        &\underset{\theta}{\min} \Bigg\{ F^{CVaR}(\theta) := \bar{q}_{\alpha \omega} \Big[ F(\theta,\omega) \Big] \notag \\ 
        &= \mathbb{E} \Big[ F(\theta,\omega) \Big| F(\theta,\omega) > q_{\alpha}(F(\theta,\omega)) \Big] \Bigg\}. \label{eq:CVAR_tuning_problem}  
    \end{align}
    
\noindent \textbf{Energy security}: 
    Parameter $\theta$ is tuned such that \eqref{eq:DLA_model} minimizes $\bpoe$ of unserved energy beyond $\zeta$:
    \begin{align}
        &\underset{\theta}{\min} \Bigg\{ F^{BPoE}(\theta)  := \bar{p}_{\zeta,\omega} \Big[R(\theta,\omega) \Big] \notag \\
        &=\!\mathbb{P} \Big[R(\theta,\omega) > z \Big| \mathbb{E} \big[R(\theta,\omega) \big| R(\theta,\omega) > z \big]\!=\!\zeta \Big]\!\Bigg\}, \label{eq:BPOE_tuning_problem} 
    \end{align} 
    where
    \begin{align}
        R(\theta,\omega)\!=\!\mathbb{E} \Big[ \sum^T_{t=0}\!L_t(S_t(\omega), X^{\rm D-LA}_t(S_t(\omega) \big| \theta)) \Big| S_0 \Big].
        \label{RA_expected_function}
    \end{align}

The resulting parameter tuning problems\eqref{eq:EX_tuning_problem}, \eqref{eq:CVAR_tuning_problem}, and \eqref{eq:BPOE_tuning_problem}
are possibly non-convex and non-smooth in $\theta$, making the optimization problem hard to solve. 
We use the stochastic gradient descent approach proposed in \cite{Ghadimi2022} and shown in Algorithm~\ref{alg:cap} to iteratively find close-to-optimal values for $\theta$.

 \begin{algorithm}
 \caption{Parameter tuning adapted from \cite{Ghadimi2022}}
 \label{alg:cap}
 \begin{algorithmic}[1]
     \State {\bf given}  $N$, $\theta^0 \in \mathbb{R}^d$, $\bar{G}^0 = 0$, sequences: $\{\eta_k\}_{k \geq 1} , \{\psi_k\}_{k \geq 1}, \{\phi_k\}_{k \geq 1} \in (0,1)$, $P_R$ 
     \State Draw random $R$ from $P_R(\cdot)$ 
     \For{$k=1,...,R$} 
        \State $\theta^k_y \leftarrow \theta^{k-1} - \psi_k \bar{G}^{k-1}$ 
        \State $\theta^k \leftarrow (1-\phi_{k}) \theta^{k-1} + \phi_{k} \theta^k_y$ \Comment{\textit{update parameter}}
        \State Generate $m_k$ scenarios $\omega^{ki}$:
        \State $G_{\eta_k}(\theta^k,\omega^{ki}) \leftarrow \frac{1}{m_k} \sum^{m_k}_{1} \frac{F(\theta^k + \eta_k \upsilon^k, \omega^{ki}) - F(\theta_k,\omega^{ki})}{\eta_k} \upsilon^k$
        \State $\bar{G}^k \leftarrow (1 - \phi_k)\bar{G}^{k-1} + \phi_k G_{\eta_k}(\theta^k,\omega^{ki})$ \Comment{\textit{update gradient}}
    \EndFor 
        
 \end{algorithmic}%
 \end{algorithm}%
\noindent
Besides an initial value $\theta_0$ and an initial gradient $\bar{G}^0$, Algorithm~\ref{alg:cap} requires hyperparameters $\{\eta_k\}_{k \geq 1}$, $\{\psi_k\}_{k \geq 1}$, $\{\phi_k\}_{k \geq 1}$ $\in (0,1)$ that define gradient smoothing and learning rates. Iteration limit $R$ is drawn from a predefined distribution $P_R$.

\section{Case Study with numerical results} \label{sec:numerical_results}

We investigate the policies derived in Section~\ref{sec:policy} on a stylized case study of the energy system model discussed in Section~\ref{sec:energy_storage_model}.
We assume daily decisions such that each time step $t$ represents a day and $T = 365$. 
We use real load profiles from the \mbox{ENTSO-E} transparency platform for Denmark's bidding zone DK2 \cite{ENTSOEDATA}. 
The peak load over the considered year is $\bar{D}=1913\,MW$, which we use to dimension other parameters shown in Table~\ref{tab:operation_parameters}.
We set the interval for hydrogen acquisition to every 7 days starting with $t=1$.
We use monthly historical gas prices in Denmark from \cite{energienet} as a proxy for hydrogen prices. 
Cost of unserved load $C^P$ is set to 1000 \$/MW and of wind curtailment penalty $C^W$ to 800 €/MW.
We implement all models in Julia with the Gurobi 10.0.2 solver. All computations have been performed on a Macbook with Apple M1 chip and 8GB RAM.

\begin{table}[]
\centering
\caption{Wind farm, battery/hydrogen storage, fuel cell parameters}
\label{tab:operation_parameters}
\begin{tabular}{@{}lllll@{}}
\toprule
Wind farm       & Initial power output         & $E_{0}$                & $0.8\cdot \bar{D}$ & MW \\ \midrule
Battery storage & Storage capacity             & $R^{\bar{E}}$ & $4\cdot \bar{D}$   & MW \\
                & Initial state of charge      & $R_0$     & $\frac{1}{2} \cdot R^{\bar{E}}$   & MW \\
                & Charging efficiency          & $\beta^c$                    & $98$                             & \% \\
                & Discharging efficiency       & $\beta^d$                    & $98$                             & \% \\
                & Max. charging capacity    & $\gamma^c$                  & $\bar{D}$     & MW \\
                & Max. discharging capacity & $\gamma^d$                   & $\bar{D}$     & MW \\ \midrule
Hydrogen storage     & Storage capacity             & $R^{\bar{H}}$ & $\frac{6}{\beta^g} \cdot \bar{D}$   & MW \\ 
                & Initial state of charge      & $R^H_0$     & $\frac{1}{2} \cdot R^{\bar{H}}$  & MW \\  \midrule
Fuel Cell & Max. discharging capacity & $\gamma^g$                   & $\bar{D}$     & MW \\
                & Fuel cell efficiency       & $\beta^g$ & $60$   & \% \\
                \bottomrule
\end{tabular}
\end{table}

We assume that the operator has access to wind power forecasts. 
Actual wind power injection is driven by atmospheric phenomena that are well-represented by persistence models (i.e., assuming only small changes between time steps) for short forecast lead times \cite{pinson2012very}. 
For longer lead times forecast accuracy decreases.
In our case study, we capture this through a martingale model of forecast evolution \cite{Sapra2004}:
\begin{align}
    &f^{E}_{t+1} = f^{E}_{t} + \epsilon_{t+1} && \forall t = 0,...,H-1.   \label{eq_rolling_forecast}
\end{align}
where $f^E_{0}$ is a given initial value.
We define the error term as $\epsilon_{t+1} \sim \mathcal{N}(0,\,\sigma_{\epsilon,t}^{2})$ with $\sigma_{\epsilon,t} = \rho_{E} f^{E}_{t}$ where $\rho_E=0.1$ is a predefined parameter.

\subsection{Constant parameter tuning}
\label{ssec:results_risk_neutral-risk_averse}

We first compare S-LA \eqref{eq:Stochastic_Look_Ahead_model}, and S-CVaR \eqref{eq:CVAR_Stochastic_Look_Ahead_model} the to D-LA \eqref{eq:DLA_model} with a constant discounting parameter tuned to reduce expected cost described in \eqref{eq:EX_tuning_problem} and no discount parameter, which is equalivalent to $\theta = 1$.
We use 100 forecast scenarios 
in each decision-making time step and evaluate the decision performance over 1000 out-of-sample scenarios.
We itemize the results in Table~\ref{table:risk-neutral_to_risk-averse}.
Using a grid search, a constant parameter of $\theta = 0.2$ achieves minimal expected cost across 1000 training scenarios. 
Any $\theta < 1$ indicates that wind forecast should be under-estimated during decision time to improve expected cost in the long run. 
Notably, the deterministic look-ahead policy significantly reduces computational time at only a 0.3\% average cost increase.

\begin{table}[]
\centering
\caption{Out-of-sample cost statistics [M\$] and average solve time.}
\label{table:risk-neutral_to_risk-averse}
\begin{tabular}{@{}l|lllll@{}}
\toprule
& Mean & $q_{80\%}$ & $q_{90\%}$ & $q_{95\%}$ & Avg. time  \\ \midrule
 $X^{D-LA}(\theta = 1)$      & 128.2  & 200.9 & 221.1 & 231.5 & \textbf{0.26s} \\
$X^{D-LA}(\theta = 0.2)$ & 121.4 & \textbf{193.2} & \textbf{215.3} & \textbf{226.1} & \textbf{0.26s} \\
$X^{S-LA}$   &  \textbf{121.0} & 194.7 & 217.7 & 227.7 & 82.88s \\
$X^{S-CVaR}$   &  121.3 & 195.1 & 217.4 & 227.7 & 96.69s  \\ 
\end{tabular}
\end{table}

\subsection{Look-up table parameter modification}
\label{ssec:results_parameter_modification}

For the look-up table tuning of $\theta$ we use Algorithm~\ref{alg:cap}
with mini-batches size $m_k = 10$ and iteration limit $N = 2000$. 
We refer to \cite{Ghadimi2022} for notes on setting hyperparameters $\{\eta_k\}, \{\psi_k\}, \{\phi_k\}$.
Table~\ref{tab:BPoE_CVaR_vs_EX_theta_comp} presents the resulting $\theta$ that reduce expected cost as in \eqref{eq:EX_tuning_problem}, 90\%-$\cvar$ as in \eqref{eq:CVAR_tuning_problem}, and $\bpoe$ with threshold $\zeta = 7000$ as in \eqref{eq:BPOE_tuning_problem}.
Each look-up table formulation performs best to its given goal as seen in Tab.~\ref{tab:BPoE_CVaR_vs_EX_theta_results_comp}.

\begin{table}[]
\centering
\setlength{\tabcolsep}{4pt} 
\caption{Look-up table for $\theta_{\tau}$ for different performance targets.}
\label{tab:BPoE_CVaR_vs_EX_theta_comp}
\begin{tabular}{c|ccccccc}
\toprule
          & \multicolumn{6}{c}{Look-ahead period}   &             \\
$\tau$  & 1     & 2     & 3     & 4     & 5     & 6     & 7 \\ \midrule
Exp. Cost        & 2.976 & 0.000 & 0.717 & 0.000 & 0.082 & 0.000 & 1.305 \\
$\bar{q}_{90\%}$ & 1.698 & 0.214 & 0.180 & 0.085 & 0.101 & 0.288 & 0.648 \\
$\bar{p}_{7000}$ & 1.079 & 0.000 & 0.000 & 0.000 & 1.319 & 0.267 & 0.000  \\ 
\end{tabular}
\end{table}

\begin{table}[]
\centering
\caption{Results of look-up table formulation of $\theta$}
\label{tab:BPoE_CVaR_vs_EX_theta_results_comp}
\begin{tabular}{@{}c|cccc|c@{}}
\toprule
            & \multicolumn{4}{c|}{Objective value [M\$]}     &                            \\
$\theta$    & Mean           & $q_{80\%}$     & $q_{90\%}$     & $q_{95\%}$   &   $\bar{p}_{\zeta}(\sum_{t=0}^TL_t)$   \\ \midrule
EC          & \textbf{120.1} & \textbf{192.5}          & 215.8          & 226.3  & 20.00 \%         \\
$\bar{q}_{90\%}$ & 123.4 & \textbf{192.5} & \textbf{214.7} & \textbf{226.1}  & 7.47 \%         \\
$\bar{p}_{7000}$ & 124.7          & 194.7          & 216.2          & 227.2  & \textbf{7.09\%} \\ 
\end{tabular}
\end{table}

\subsection{Influence of $\theta$ in reducing energy shortfalls}
\label{ssec:results_bpoe}

\begin{figure}
    \centering        
    \includegraphics[width=0.9\linewidth, keepaspectratio]{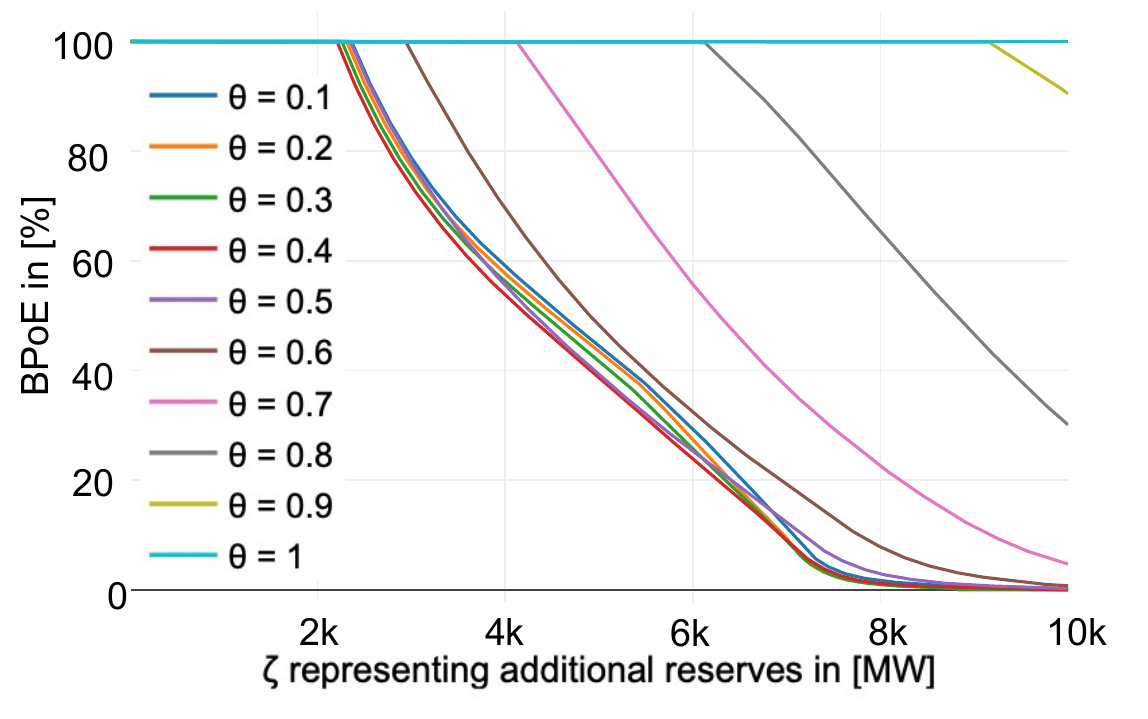}
        \caption{Influence of a constant $\theta$ on the $\bpoe$ of shortfall events.}
        \label{fig:BPOE_constant_parameter_01}
\end{figure}

Finally, we investigate how a constant parameter $\theta$ in $X_t^{\rm D-LA}(\theta)$ impacts the resulting $\bpoe$ of a given security threshold $\zeta$.
Fig.~\ref{fig:BPOE_constant_parameter_01} shows the $\bpoe$, as per \eqref{Loss_function}, for 10 values of $\theta$ over the 1000 out-of-sample runs. 
For $\theta=1$, i.e., operating the system under the assumption that the given forecast is the true wind-power injection, any of the given thresholds are exceeded with $\bpoe=1$. With higher conservatism ($\theta < 1$) $\bpoe$ is reduced depending on the target threshold $\zeta$.

\section{Conclusion} \label{sec:conclusion}

This paper derives a risk-aware linear policy approximation for energy-limited and stochastic energy systems with rolling forecasts. 
In particular, the inclusion of the BPoE in the tuning of the parameter enables the system operator to certify that the implemented operation policy can achieve predefined reliability targets.
There are two main avenues for future work. On the one hand, we will increase the model fidelity through other parameter modifications and by including several parameter modifications across several sources of uncertainty.
On the other hand, we will investigate how the gained insights on parameter $\theta$ can be used as a means to discount wind and solar time series in deterministic investment models such that they avoid over- or under-estimation of available energy.

\bibliographystyle{IEEEtran}
\bibliography{bibliography.bib}

\end{document}